\documentstyle[epsfig,epsf]{article}
\begin{document}

\begin{center}
{\Large {\bf Nuclear Track Detectors for Particle Searches}}
\end{center}

\vskip .7 cm

\begin{center}
S. Manzoor$^{1,2}$, 
S. Balestra$^1$,  
M. Cozzi$^1$, 
M. Errico$^1$, 
G. Giacomelli$^1$,  
M. Giorgini$^1$,  
A. Kumar$^{1,3}$,     
A. Margiotta$^1$,  
E. Medinaceli$^1$,  
L. Patrizii$^1$,  
V. Popa$^{1,4}$,  
I.E. Qureshi$^2$, 
and V. Togo$^1$
 \par~\par

{\it  1. Phys. Dept. of the University of Bologna and INFN, Sezione di 
Bologna, Viale C. Berti Pichat 6/2, I-40127 Bologna, Italy \\ 
2. PRD, PINSTECH, P.O. Nilore, Islamabad, Pakistan \\
3. Dept. Of Physics, Sant Longowal Institute of Eng. and Tech., Longowal 
148 106 India \\
4. Institute of Space Sciences, Bucharest R-77125, Romania} 

\par~\par
{\small Presented at the 10$^{th}$ Topical Seminar on Innovative Particle and Radiation 
Detectors, 1-5 October 2006, Siena, Italy.}

\vskip .7 cm
{\large \bf Abstract}\par
\end{center}

{\normalsize In this paper we report a search for intermediate mass magnetic 
monopoles and nuclearites using CR39 and Makrofol Nuclear Track Detectors 
(NTDs) of the SLIM large area experiment, 440 $m^2$ exposed at the high 
altitude laboratory of Chacaltaya (Bolivia) and about 100 
$m^2$ at Koksil, Himalaya (Pakistan). We discuss the new chemical etching and 
improved analysis of the 
SLIM CR39 sheets. Preliminary limits are based on 316 $m^2$ of CR39 NTDs 
exposed 
for 3.9 y.  }

\vspace{5mm}

\large
\section{Introduction}\label{sec:intro}Grand Unified Theories (GUT) of 
the strong and electroweak interactions predict the existence of magnetic 
monopoles (MMs) of large mass ($10^{16}$ - $10^{18}$ GeV/$c^2$), produced in 
the early 
Universe. The MACRO experiment has set the best limits on GUT MMs for 
$4 \times 10^{-5} < \beta < 1,~ \beta $= v/c [1-2].\par
Intermediate Mass Magnetic Monopoles (IMMs) may have been produced in 
later phase transitions in the Early Universe; IMMs with masses 
$10^7 < m_M < 10^{13}$ GeV and $g > g_D$ could be present in the cosmic 
radiation 
and could be accelerated to relativistic velocities in one coherent 
domain of the galactic magnetic field [3-4]. Thus one may look for 
$\beta \geq$ 0.1, 
heavily ionizing IMMs.\par
We used CR39 nuclear track detectors (NTDs) for a variety of studies: 
searches for magnetic monopoles and nuclearites, determination of 
fragmentation cross-sections, search for nuclear fragments with fractional 
charges, measurement of the composition of primary cosmic rays [5-6].\par
The main aim of the present work is the determination of the optimal etching 
conditions to achieve the best surface quality and to reduce the umber of 
background tracks in CR39 and Makrofol NTDs used in the SLIM experiment.

\section{Experimental}The SLIM experiment planned for the search of IMMs and Strange Quark Matter (SQM) is based on 440 $m^2$ of NTDs installed at the Chacaltaya high altitude lab. 
(5230 m a.s.l.) since 2001 [7]. Another 100 $m^2$ of NTDs were installed at Koksil, Himalaya 
(Pakistan, 4270 m a.s.l.) since 2002. The radon activity and the flux of cosmic ray neutrons 
were measured at Chacaltaya [8]. \par
The SLIM basic unit is a ``wagon" composed of 3 layers of 1.4 mm thick 
CR39, 3 layers of 500 
$\mu m$ Makrofol, 2 layers of 250 $\mu m$ lexan sheets and an aluminium absorber of 1 mm thick, 
see Fig. 1. \par
Extensive test studies were made in order to improve the etching procedures of CR39 and Makrofol
 NTDs, and to improve the scanning efficiency and analysis procedures. In this note we discuss 
the etching procedures for the SLIM CR39 and Makrofol NTDs. \par
The CR39 foils from SLIM modules exposed to cosmic rays and to 1 A GeV $Fe^{26+}$ ions were 
originally etched without alcohol. We found several background tracks of 10-17 $\mu m$ 
range due to carbon, oxygen and proton recoils produced in the interactions of ambient neutrons; see 
Figure 2(a, b): the surface quality of both sheets was poor. In these conditions, it would be 
difficult to scan the detectors. In order to improve the surface quality and to eliminate the recoil 
tracks we etched the CR39 sheets with 8N KOH + 1.5$\%$ ethyl alcohol at 75 $^\circ C$ for 30 hr 
(see Figure 2(c, d)). The surface quality of the etched SLIM CR39 sheets improved and most of the 
recoil tracks were removed. Moreover the detector is transparent and scanning is easier. The tracks 
of the relativistic iron ions and their fragments have sharp contours and can be easily measured with the automatic image analyzer system "ELBEK". The threshold of the detector increased to 
Z/$\beta \sim$ 17.

\begin{figure}[!ht]
\begin{center}
\mbox{\epsfig{figure=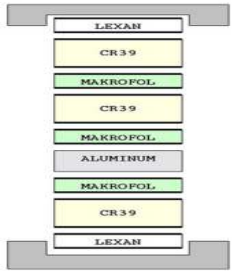,height=4 cm,width=4 cm}}
\caption{Sketch of a SLIM "wagon", sealed in an aluminium plastic bag. }
\label{fig:1}
\end{center}
\end{figure}

CR39 ``strong etching conditions" are now used for the first (top) CR39 sheet in each SLIM module 
(see Fig. 1), in order to produce "large tracks", easier to detect during scanning. ``Soft etching 
conditions" (6N NaOH + 1$\%$ Ethyl alcohol at 70 $^\circ C$ for 40 hours) are applied to the other 
CR39 layers in a module [9], if a candidate track is found in the first layer. Soft etching allows 
more reliable measurements of the restricted energy loss (REL) and of the direction of the incident 
particle. \par
The Makrofol detectors are etched in 6N KOH + 20$\%$ ethyl alcohol at 50 $^\circ C$ [10-11].

\section{Calibration}The CR39 and Makrofol detectors were calibrated with 158 A GeV 
 $In^{49+}$ and $Pb^{82+}$ ions at the CERN SPS and 1 A GeV $Fe^{26+}$ ions at the BNL AGS [11]. 
The base areas of the ``post-etched cones" were measured with an automatic image analyzer system. 
Fig. 3 shows the average base area distribution of 158 A GeV $In^{49+}$ ions and their fragments, 
the averages were made on two front faces of CR39 sheets. With the above mentioned etching, the peaks 
are well separated from Z/$\beta$ = 7 to 45 and charge can be assigned to each individual peak. \par
The ``strong" etching conditions for CR39 allow the detection of MMs with one unit Dirac charge 
(g=$g_D$), for $\beta$ around $10^{-4}$ and for $\beta > 10^{-2}$, the whole $\beta$-range of 
4 $\times 10^{-5} < \beta < 1$ for MMs with g $\geq 2 ~ g_D$, for dyons and for nuclearites. 
The Makrofol is useful for the search for fast MMs. CR39 and Makrofol can detect nuclearites with 
$\beta \geq 10^{-3}$.

\begin{figure}[ht]
\begin{center}
\mbox{\epsfig{figure=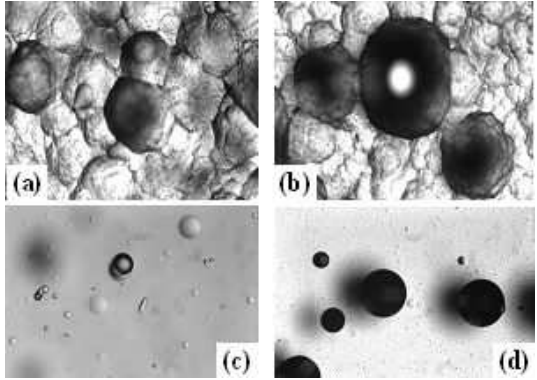,height=5 cm}}
\caption{(a) SLIM CR39 sheet, (b) tracks of 1 A GeV $Fe^{26+}$ ions and their fragments in CR39 using 
8N NaOH 90 $^\circ C$ for 48h without alcohol, (c) SLIM CR39 sheet and (d) the tracks of 1 A GeV 
$Fe^{26+}$ ions and their fragments in CR39 with ``strong" etching with 1.5$\%$ ethyl alcohol at 75 
$^\circ C$, ($G_{tot.}$ = 20x).}
\label{fig:2}
\end{center}
\end{figure}

\begin{figure}[ht]
\begin{center}
\mbox{\epsfig{figure=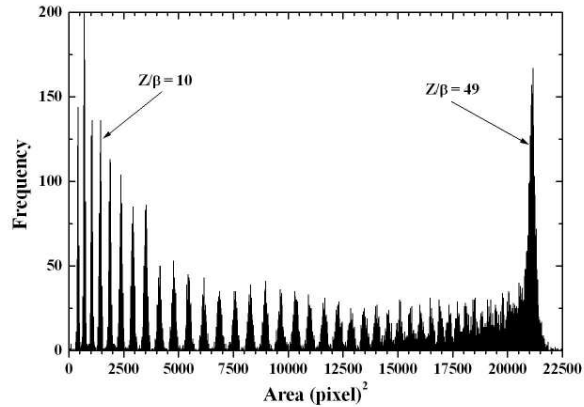,height=6 cm}}
\caption{Base area distribution of etched cones in CR39 from 158 A GeV $In^{49+}$ ions and their 
fragments (averages of 2 front face measurements)}
\label{fig:3}
\end{center}
\end{figure}

\section{Results and Discussion}The top CR39 layer (see Fig. 1) of each wagon 
was chemically treated 
with our improved ``strong" etching conditions with alcohol to reduce the 
thickness of CR39 sheets 
from 1400 $\mu m$ to 900 $\mu m$ and to make large tracks. The etched CR39 
sheets were transparent 
and with low background. The sheets were scanned with stereomicroscopes 
searching for passing tracks, 
which form a double cone, or through holes. \par
In order to compute the p-values and incident angles $\theta$ for the front 
and backsides, 
the track major and minor axes are also measured. Finally, a track is defined 
as a ``candidate" if p 
and $\theta$ on the front and backsides are equal to within 15$\%$. In the 
presence of a candidate, 
the lowermost CR39 layer was etched in ``soft conditions", and an accurate 
scan was performed with an 
optical microscope at high magnification. Up to now no two-fold coincidences 
or through holes were 
observed.

\section{Conclusions}We analyzed an area of 316 $m^2$ of SLIM CR39 sheets 
exposed for 3.9y at the Chacaltaya high altitude lab. No candidate was observed; the 90$\%$ CL upper flux limits for downgoing IMMs with g = $g_D$, 2$g_D$, 
3$g_D$ and M+p are plotted in Fig. 4 versus $\beta$.

\begin{figure}
\begin{center}
\vspace{-1.5cm}
\mbox{\epsfig{figure=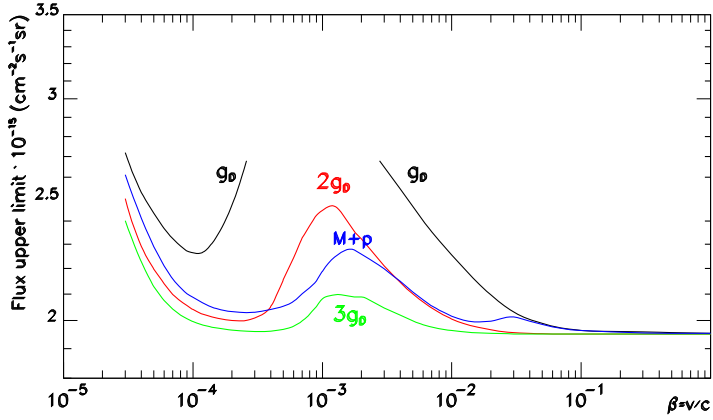,height=7 cm}}
\caption{90$\%$ C.L. flux upper limits for downgoing intermediate mass 
magnetic monopoles with g = $g_D$,  2$g_D$, 3$g_D$ and M+p.}
\label{fig:4}
\end{center}
\end{figure}

Intermediate mass nuclearites lose more energy than g = $g_D$ magnetic 
monopoles; thus the limit is at the level of the MM limit for 
$\beta > 0.1$ ($\Phi < 1.89 \times ~ 10^{-15} ~ cm^2 ~ s^{-1} ~ sr^{-1}$). 
A special search was made for very light nuclearites; since we found no 
candidate the limit is as indicated for IMMs at high $\beta$.

{\bf Acknowledgements.} We thank many colleagues for cooperation and 
discussions. We acknowledge the contributions of our technical staff in 
Bologna, and the Chacaltaya High Altitude Laboratory. We thank INFN and ICTP 
for providing fellowships and grants to non-Italian citizens.

\end{document}